\documentstyle[11pt,moriond,epsfig]{article}

\def\Journal#1#2#3#4{{#1} {\bf #2}, #3 (#4)}

\def\NIMA{{\em Nucl. Instrum. Methods} A}

\def\PLB{{\em Phys. Lett.}  B}
\def\PRL{\em Phys. Rev. Lett.}
\def\PRD{{\em Phys. Rev.} D}

\def\be{\begin{equation}}
\def\ee{\end{equation}}
\def\bea{\begin{eqnarray}}
\def\eea{\end{eqnarray}}

\begin{document}
\vspace*{4cm}
\title{RESULT OF T-VIOLATING MUON POLARIZATION MEASUREMENT 
       IN THE 	 $K^+\rightarrow \pi^0 \mu^+ \nu$ DECAY}

\author{J. IMAZATO \\
         ( for the KEK E246 collaboration$^*$
)} 

\address{Institute of Particle and Nuclear Physics, KEK, Oho 1-1 \\
Tsukuba-shi, Ibaraki-ken, 305-0801 Japan}

\maketitle

\abstracts{
A search for T-violating muon polarization in the $K^+ \rightarrow \pi^0 \mu^+ \nu$ 
decay has been performed at KEK using stopped kaons. A new improved limit was obtained, 
$P_T= -0.0017 \pm 0.0023 (stat) \pm 0.0011 (syst)$, corresponding to a  
90 \% confidence limit of $\vert P_T\vert < 0.0050$. The T-violating parameter was
also determined to be Im$\xi = -0.0053 \pm 0.0071 (stat) \pm 0.0036 (syst)$, and
$\vert{\rm Im}\xi\vert < 0.016$ (90 \% CL). 
}

\section{Transverse Muon Polarization}

Experiment E246 \footnote[1]{
The E246 Collaboration:
 M.~Abe, K.~Horie, Y.~Igarashi, J.~Imazato, G.Y.~Lim, T.~Yokoi ({\it KEK});
 M.~Aliev, V.~Anisimovsky, A.P.~Ivashkin, M.M.~Khabibullin,  A.N.~Khotjantsev, 
 Yu.G.~Kudenko, A.~Levchenko, O.V.~Mineev, N.~Okorokova, N.~Yershov ({\it INR, 
  Russian Academy of Sciences});
 M.~Aoki, Y.~Kuno,  S.~Shimizu ({\it Osaka Univ.}); 
 Y.~Asano ({\it Univ.of Tsukuba}); 
 T.~Baker, C.~Rangacharyulu, Y.-M.~Shin ({\it Univ.of Saskatchewan}); 
 M.~Blecher ({\it Virginia Polytechnic Institute and State Univ.});  
 P.~Depommier ({\it Univ.de Montr$\acute{e}$al});  
 M.~Hasinoff ({\it Univ.of British Columbia}); 
 K.S.~Lee, K.S.~Sim ({\it Korea Univ.}); 
 J.A.~Macdonald ({\it TRIUMF});  
 Y.-H.~Shin ({\it Yonsei Univ.}).} 	
 has searched for a violation of time-reversal invariance (T-violation) 
by means of a precise measurement of the transverse muon polarization, $P_T$, 
in $K^+ \rightarrow \pi^0 \mu^+ \nu$ decay ($K^+_{\mu3}$) at KEK. This polarization 
is defined as the component of the muon polarization perpendicular to the decay plane, 
namely $P_T={\vec{s}_{\mu}\cdot (\vec{p}_{\pi}\times \vec{p}_{\mu}) / |\vec{p}_{\pi} 
\times \vec{p}_{\mu}| }$, and its non-zero value is a clear signature of T-violation 
with T-odd character because of the negligible level of any spurious final-state 
interaction effects \cite{fsi}of less than $10^{-5}$. T-violation is itself an important 
symmetry violation to be tested, since, at the same time, it provides knowledge of CP-
violation through the CPT theorem. The important feature in the present case is the fact 
that the contribution from  the standard model (SM) Kobayashi-Maskawa scheme is negligibly 
small \cite{big} ($\sim 10^{-7}$). Thus, an observation of $P_T$ above the level of 
$10^{-5}$ uniquely signifies the discovery of a CP violation mechanism other than the SM. 
Several theoretical models \cite{mod},eg, three-Higgs-doublet models, leptoquark models, 
and  some class of supersymmetric models with R-parity violation or squark mixing, have 
been considered and they can produce $P_T$ as large as $10^{-3}$ without conflicting with 
other experimental constraints. The $K_{\mu3}$ $P_T$ measurement has a long history both 
in $K_L$ and $K^+$ decays including the most recent $K^+$ at BNL-AGS about 20 years 
ago \cite{bla}. All of them, however, ended with upper limits. E246 aimed for improving 
the limit further in statistical accuracy as well as in the systematic errors. The final 
result of this experiment has recently been obtained, and it is presented in this talk.  
       
\begin{figure}[t]
\psfig{figure=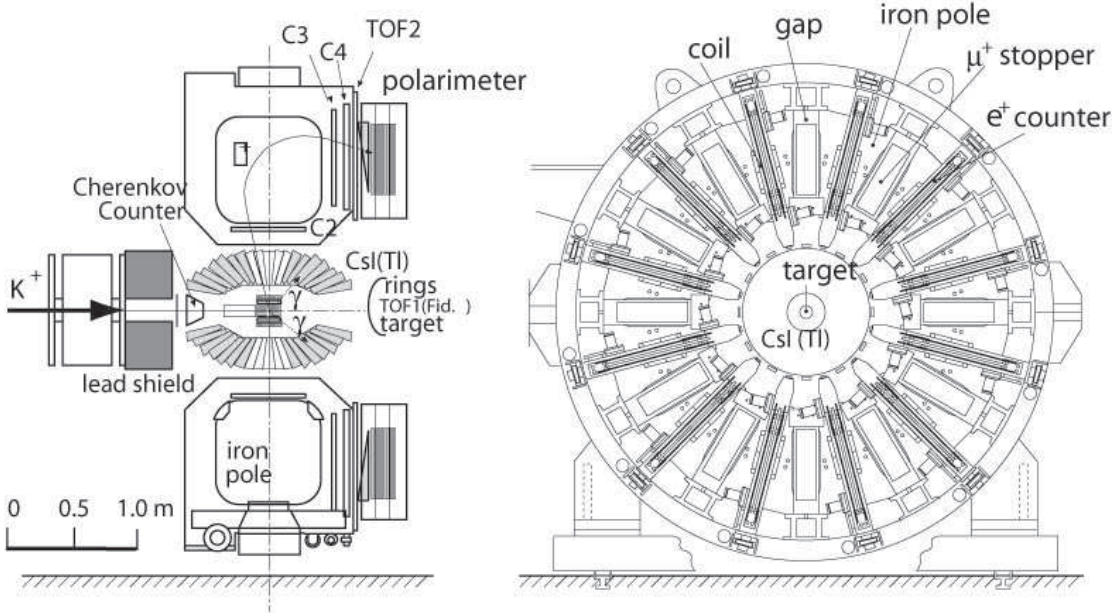, width=15cm}
\caption{E246 experiment setup. End view (right) and
         cross section side view (left).}
\label{fig:setup}
\end{figure}

\section{E246 Experiment}

\subsection{Progress}

The main feature of the E246 experiment is the use of kaon decays at rest in contrast 
to
all the previous experiments which used kaon decays in flight. This 
enabled a measurement of all decay kinematic directions, and, in this way, a double 
ratio
measurement with suppressed systematic errors was realized. After the detector 
construction
between 1992 to 1995, data taking was performed for 5 years from 1996 to 2000. In 1999 
we published
the first result \cite{abe} from the first 25\% of the data, giving ${\rm Im}\xi = -0.013 
\pm 0.016 (stat) \pm
0.003(syst)$, where Im$\xi$ is the T-violating physics parameter \cite{cab} in 
$K_{\mu3}$ decay. $\xi$ is
 defined as the form factor ratio, $\xi = f_{-}/f_{+}$. A complete analysis of the 
entire data set  
has been done carefully after the completion of the data acquisition. The E246 data 
also contained
several byproduct physics results, such as the
$P_T$ measurement in $K^{+}\rightarrow\mu^+\nu\gamma$ \cite{ani} and decay-form-factor 
related
physics \cite{shi}.

\begin{figure}[ht]
\psfig{figure=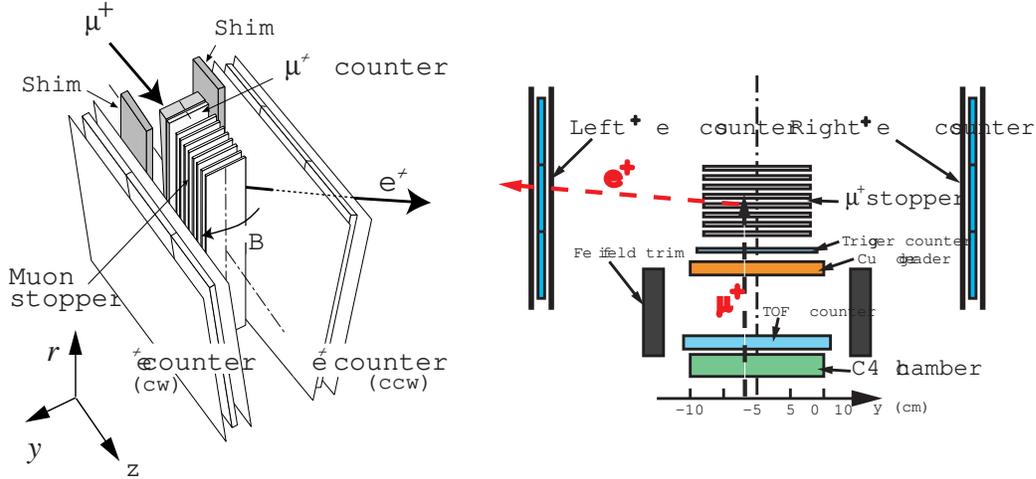, width=13cm} 
\caption{Schematic view of the muon polarimeter one sector (left) and its cross 
section (right).}
\label{fig:pol}
\end{figure}

\subsection{Experimental Setup}

The experimental setup and the principle of the experiment have been well 
described \cite{abe,jam}.
A schematic view of the setup is shown in Fig.1. The detector system consists of 
a charged particle 
tracking system, a CsI(Tl) calorimeter with 762 segmented crystals for 
$\pi^0$ detection, and a muon 
polarimeter to measure the decay positron asymmetry. Incoming kaons, triggered by a 
Cherenkov counter, were stopped
in an active fiber target where $K_{\mu3}$ decay takes place with a branching ratio of 
3.2\%. $P_T$ was 
searched for as the azimuthal ($\phi$) polarization  ($y$ component in Fig.2) of 
$\mu^{+}$ emitted 
radially (in the $r$ direction) and stopped in the pure Al stoppers when a $\pi^{0}$ 
was tagged in the 
forward ($fwd$) or the backward ($bwd$) direction relative to the detector ($z$) axis; 
events from 
$fwd$ and $bwd$ $\pi^{0}$s have $P_T$ with opposite signs. 

\subsection{Muon Polarimeter}

A schematic view of the polarimeter and its cross section are shown in Fig.2.
Muons entering the polarimeter are slowed down through a degrader and stopped in stack 
of 
pure Al plates, with spacing in-between to reduce scattering and absorption of decay 
positrons. A flux 
of magnetic field was guided by iron plates to apply a field on the stopper with 
strength
of 200-300 Gauss to hold the azimuthal component while rotating the other components. 
There is neither spin relaxation nor initial loss in the pure Al. The azimuthal 
polarization was 
measured as an asymmetry 
$A=[N_{cw}-N_{ccw}]/[N_{cw}+N_{ccw}] \sim (N_{cw}/N_{ccw} -1)/2$ 
between clockwise ($cw$) and counter-clockwise ($ccw$) $e^{+}$ counts, $N_{cw}$ and 
$N_{ccw}$.

\begin{figure}[t]
\psfig{figure=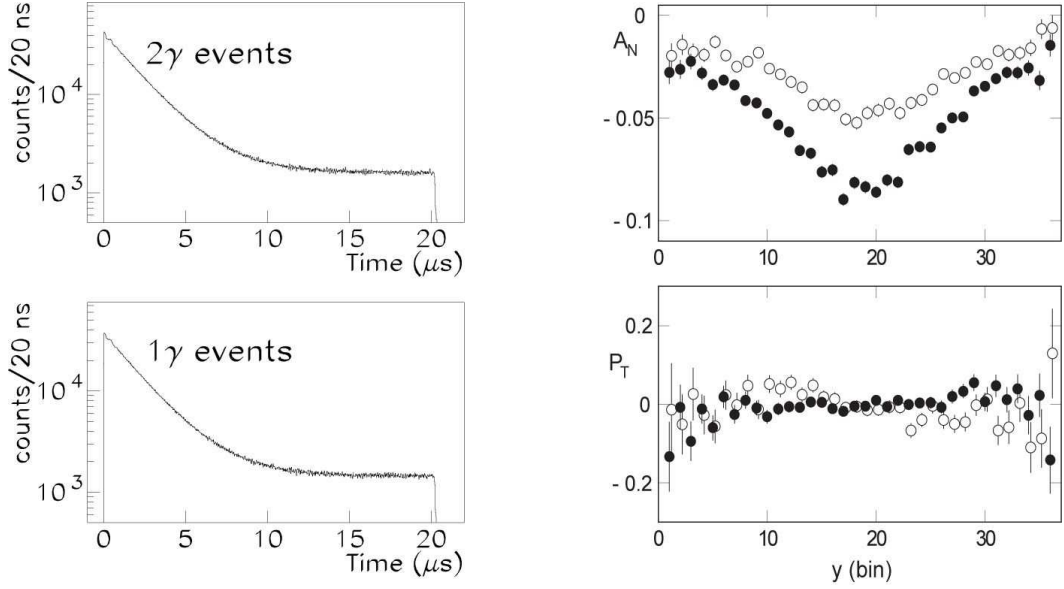, width=14cm}
\caption{Time spectra of decay positrons (left) and result of $P_T$ as a function 
polarimeter axis $y$
         (right). In the upper part the asymmetry $A_N$ which represents the 
analyzing power is also plotted.
          Black dots ($\bullet$) are 2$\gamma$ events and open circles ($\circ$) are 
1$\gamma$ events.}
\label{fig:spec}
\end{figure}

\section{Analysis}

\subsection{$K_{\mu3}$ Event Selection}

$K^+_{\mu3}$ events were selected in terms of 1) the charged-particle momentum 
$p_{\mu}$,
2) the charged particle mass from the time-of-flight measurement $m_{\mu}$, and 3) 
the CsI(Tl)
information on the $\pi^0$ ( $m_{\gamma\gamma}$ for 2-photon events and $E_{\gamma}$ for 
one-
photon events: the CsI(Tl) has 12 holes for muons to pass into the spectrometer as 
well as 
two beam in/out holes.  $\pi^0$'s were identified not only with 2 photons but also 
as one photon 
with relatively high energy). Necessary conditions for the active kaon target, the 
kinematics, and 
veto counters {\it etc.} were also imposed. The most significant background was from
$\pi^+$ in-flight decay muons, but this could be suppressed down to a level less 
than several \% after the tracking quality cuts such as the fit $\chi^2$.
Thanks to the very good timing performance of the CsI(Tl) crystals, the accidental 
backgrounds in the calorimeter could be very efficiently suppressed. 

\subsection{Two-Analysis Method}

We employed the two-analysis method, namely two completely 
independent analyses, A1 and A2, pursued their own best event selection conditions 
with their own analysis criteria. The basic selections of good $K^{+}_{\mu3}$ events  
were almost common, but details of 1) charged particle tracking, 2) CsI(Tl) 
clustering, cut variables and the cut points were generally different. This method 
provided 
the means of a data quality cross-check of the selected events. All the selected 
events were then sorted into common ($A1\cdot A2$) events and two sets of uncommon events 
($\overline{A1}
\cdot A2$ and $A1\cdot\overline{A2}$) separately for 2$\gamma$ and 1$\gamma$, 
thus providing 6 final data sets. 
Slight differences between the two analyses led to a non-negligible amount of uncommon 
good events 
in each analysis. The maximum sensitivity to $P_T$ is provided by the $fwd$ and 
$bwd$ regions 
of $\pi^{0}$ (2$\gamma$) or photons (1$\gamma$) with $\vert \cos 
\theta_{\pi^0(\gamma)}\vert >0.342$, 
where $\theta_{\pi^0(\gamma)}$ is the polar angle, and  $A$ for $fwd$ and $bwd$ were 
calculated
as $A_{fwd}$ and $A_{bwd}$, respectively..

\begin{figure}[t]
\psfig{figure=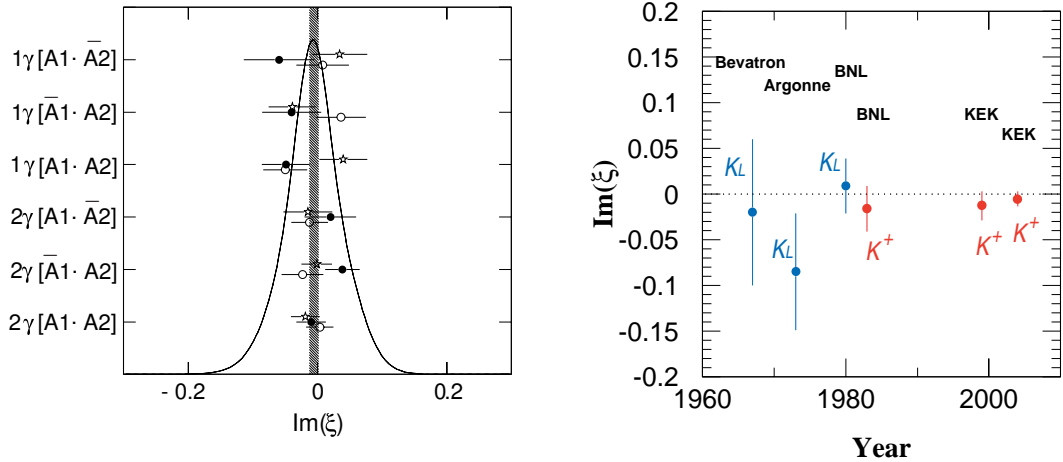, width=14cm}
\caption{Ideogram of Im$\xi$  for 18 data sets from 1996~1997 ($\bullet$), 1998 
($\circ$) and 
         1999~2000 ($\star$) (left) and the history of previous $P_T$ limits 
         and our E246 result (right).}
\label{fig:res}
\end{figure}

\subsection{Data Quality Check}

The data were grouped into three experimental 
periods of (I) 1996-1997, (II) 1998, and (III) 1999-2000, each having nearly the same 
beam conditions and 
almost the same amount of data, giving 3 $\times$ 6 = 18 data sets. First, the null 
asymmetry $A_0=(A_{fwd}+A_{bwd})/2$
was confirmed to be close to zero for each data set. Next, the polarimeter 
sensitivity was 
checked by means of the asymmetry associated with the large in-plane polarization 
$P_N$. The selected events were
rearranged into ``left'' and ``right'' categories instead of $fwd$ and $bwd$ and the 
asymmetry
$A_N=(A_{left}-A_{right})/2$ was calculated. The quantity $A_N/P_N 
<\cos\theta>$ represents the
sensitivity to the polarization measurement, where $<\cos\theta>$ is the kinematical 
attenuation of 
$P_N$. The third data quality check was done by means of the distribution of the decay plane 
angles relative to the polarimeter axis. The asymmetry of this distribution would induce 
an admixture of an in-plane polarization. All the data sets passing these tests were  
then used for $P_T$ extraction. 

\subsection{Polarimeter Analysis}

The muon stoppers had finite sizes in the $y$ and $r$ directions (Fig.2). In order 
to remove the intrinsic geometrical asymmetry we employed 
the distribution information from the C4 tracking chambers located in front of the stopper. 
$P_T$ for each data set was evaluated as the average of the contributions $P_T(y)$ from 
each part of the stopper from $y$ = -9.0~cm to +9.0~cm as;
\begin{equation}
    P_T = \int P_T(y) w(y) dy 
\end{equation}
where $w(y)$ is the weight and $P_T(y)=A_T(y)/[\alpha(y) <\cos\theta_{T}>]$
with the $y$-dependent asymmetry $A_T(y)$  and analyzing power $\alpha(y)$. 
$A_T(y)$  defined as 
$ A_T(y) = [(A_{fwd}(y)-A_{bwd}(y)]/2$  was  free from the intrinsic geometrical  
asymmetry and from the  
muon stopping densities, and therefore cancelled the systematic errors common for 
$fwd$/$bwd$ events. The analyzing
power $y$ dependence could  be calibrated using the positron asymmetry 
$A_N(y)$ associated with the normal 
polarization $P_N$ as $\alpha(y)\sim A_N(y) $. The absolute value of $\alpha$ was 
calibrated by a Monte Carlo simulation. 
Fig.3 shows 
$A_N(y)$, and $P_T(y)$ thus calculated which is nearly constant with only slight gradients 
for both 2$\gamma$ 
and 1$\gamma$ events. This is due to the different muon stopping distributions in the  
$r$ direction between $fwd$ and 
$bwd$ events. $P_T$ was calculated by averaging $P_T(y)$. The effect of the 
$P_T(y)$ gradients could be 
eliminated due to the symmetric nature about $y=0$ in this summation. The factor 
$<\cos\theta_{T}>$ was  
evaluated for each data set by a Monte Carlo calculation taking into account 
realistic  background conditions for each data set.

\section{Result and Systematic Errors}
The transverse polarization $P_T$ was calculated as the average of the 18 values to 
be  $P_T= 
-0.0017 \pm 0.0023 $. Thus, no T violation was observed. The conversion to the T-violating 
physics parameter 
Im$\xi$ was done using a conversion coefficients $\Phi=0.327 (0.287)$  from a Monte 
Carlo simulation for $
2\gamma$($1\gamma$).  Its ideogram is shown in Fig.4 with the average of ${\rm 
Im}\xi= -0.0053 \pm 0.0071 $.  
%

The major systematic errors are listed in Table I.
Almost all the systematics were cancelled due to the summation of the rotationally 
symmetric 12 sectors and the double ratio between $fwd$ and $bwd$ events.
The few remaining errors give rise to a small admixture of $P_N$ resulting in a spurious 
$P_T$ effect. There are some contributions from misalignments of detector
elements and the muon spin rotation field. The small shifts of the decay plane normal 
distribution, $\theta_{r}$ and $\theta_{z}$,  were treated as an error. The effect of 
muon multiple scattering through 
the Al degrader can cause a difference in the actual muon stopping distribution, and 
thus produce a spurious $A_T$
through the intrinsic geometrical asymmetry as large as $\delta P_T = 7.1 
\times 10^{-4}$. 
The total size of the systematic error was calculated as the quadratic sum of all 
the contributions resulting in $\delta P_T = 1.1 \times 10^{-3}$  which is much smaller 
than the statistical error.   
Finally, the ideogram of 18 Im$\xi$ values shows a good behavior with a fit to a 
constant with $\chi^2/\nu=0.78$.

 \begin{table}
 \begin{center}
 \caption{Summary of systematic errors.}
 \begin{tabular}{|l|c|}
 \hline
  Source  & $\delta P_T \times 10^4 $  
     \\
 \hline
  $e^+$ counter misalignment         & 2.9      \\
  Misalignments of other counters    & 2.6      \\
  Misalignment of $\vec{B}$ field    & 6.1      \\
  $K^+$ stopping distribution        &$<3.0$    \\
  Decay plane rotations              & 1.4      \\
  $\mu^+$ multiple scattering        & 7.1      \\         
  Backgrounds                        &$<2.0$    \\
  Analysis                           & 4.0      \\
 \hline                                       
 Total                               & $<11.4$  \\
 \hline
 \end{tabular}
 \end{center}
\end{table}

\section{Summary}

The KEK E246 experiment has obtained the improved limits of
\begin{eqnarray*}
       P_T= -0.0017 \pm 0.0023 (stat) \pm 0.0011 (syst)     \\
       {\rm Im}\xi = -0.0053 \pm 0.0071 (stat) \pm 0.0036 (syst),
\end{eqnarray*}    
showing no evidence for T violation. The 90\% CL's are given as $\vert 
P_T \vert <
0.0050$ and  $\vert$Im$\xi \vert < 0.016$ by adding statistical and systematic errors 
quadratically.
This limit of Im$\xi$ is a factor 3 improvement over the last BNL-AGS experiment \cite{bla}. 
Our results constrain the lightest Higgs mass and other parameters in the framework of non-SM 
models \cite{mod} better than or complementary to other constraints such as the neutron 
electric dipole moment $d_n$ and $B$ meson decays.  

\end{document}